\documentclass[twocolumn,amsmath,amssymb,prd]{revtex4}
\renewcommand{\narrowtext}{\begin{multicols}{2} \global\columnwidth20.5pc}
\def\widetext{\end{multicols} \global\columnwidth42.5pc}
\usepackage{graphicx}
\usepackage{esvect}

\def\be{\beta}
\def\ga{\gamma}

\def\th{\theta}

\def\la{\lambda}

\def\ph{\phi}

\def\ch{\chi}

\def\om{\omega}
\def\Ga{\Gamma}
\def\De{\Delta}

\def\La{\Lambda}

\def\Ps{\Psi}
\def\Om{\Omega}

\def\cl{{\cal L}}

\def\fr#1#2{{{#1} \over {#2}}}
\def\frac#1#2{\textstyle{{{#1} \over {#2}}}}

\def\prt{\partial}

\def\ket#1{|{#1}\rangle}
\def\bra#1{\langle{#1}|}

\def\half{{\textstyle{1\over 2}}}

\def\lsim{\mathrel{\rlap{\lower4pt\hbox{\hskip1pt$\sim$}}
    \raise1pt\hbox{$<$}}}
\def\gsim{\mathrel{\rlap{\lower4pt\hbox{\hskip1pt$\sim$}}
    \raise1pt\hbox{$>$}}}
    \def\ket#1{|{#1}\rangle}

\def\sqr#1#2{{\vcenter{\vbox{\hrule height.#2pt
         \hbox{\vrule width.#2pt height#1pt \kern#1pt
         \vrule width.#2pt}
         \hrule height.#2pt}}}}

\def\X{\hat X}
\def\Y{\hat Y}
\def\Z{\hat Z}
\def\E{\hat E}
\def\S{\hat S}
\def\U{\hat U}

\def\Re{\hbox{Re}\,}
\def\Im{\hbox{Im}\,}

\def\ol#1{\overline{#1}}
\newcommand{\beq}{\begin{equation}}
\newcommand{\eeq}{\end{equation}}
\newcommand{\bea}{\begin{eqnarray}}
\newcommand{\eea}{\end{eqnarray}}
\newcommand{\rf}[1]{(\ref{#1})}
\newcommand{\bit}{\begin{itemize}}
\newcommand{\eit}{\end{itemize}}

\def\prt{\partial}

\def\ol#1{\overline{#1}}

\newcounter{tc1}\newcounter{tc2}
\newcounter{tr1}\newcounter{tr2}

\begin{document}

\title{Testing CPT symmetry with correlated neutral mesons
}

\author{\'Agnes Roberts}

\affiliation{Indiana University Center for Spacetime Symmetries,
Bloomington, Indiana 47405, USA}

\date{August 2017}

\begin{abstract}
This work gives a general overview of phenomenology developed for neutral-meson searches for CPT violation in the framework of the Standard-Model Extension with focus on meson factories. It gives a comparison of notations and fundamental approach in the formalism used by the different experiments. Asymmetries and possible experimental investigations are presented for tests of the momentum-dependent phenomenological parameter of CPT violation in correlated neutral-meson oscillations. The general results apply to any mesons produced as correlated meson pairs and address the issue of decoherence as a consequence of direction dependence. An analysis is given considering kinematics and orientation of the improved Belle II experiment.  
\end{abstract} 
\maketitle
\section{Introduction}
\label{Int}
Neutral-meson oscillations are one of the classic tests of CPT symmetry, the combined symmetry of (C) charge conjugation, (P) parity and (T) time reversal. According to the CPT theorem, any local Lorentz-invariant relativistic quantum field theory of point particles is invariant under this combination. It is one of the few exact symmetries of the Standard Model; so any  violation would indicate physics beyond it. The violation of CP symmetry has been identified as a necessary physical phenomenon for the baryon-antibaryon asymmetry of the universe, but known violations do not offer sufficient explanation for it.

Being identical for strong and electromagnetic interactions but oscillating into one another by higher-order weak processes, the neutral mesons present a highly sensitive interferometric system and a rich source of tests for new physics. Due to the observed CP and T violation, special attention is paid to its spacetime-symmetry experiments. The phenomenology to conduct these searches is well established. 

CP and T violation have been observed and some significant constraints were given on CPT violation. High-sensitivity detectors studying neutral mesons, including high-energy colliders and particle-specific meson factories such as the $K$ and $B$ factories, are steadily probing the boundaries between the Standard Model and beyond, to find evidence of Planck-scale physics in quantum-gravity scenarios. 

One of these scenarios is the Standard-Model Extension (SME), a framework based on spontaneous CPT symmetry breaking. Theories with extended particles in higher dimensions have explored spontaneous symmetry breaking involving tensor fields, leading to models with possible Lorentz and CPT violation \cite{KySam89}. Discussions of spontaneous local Lorentz and diffeomorphism symmetry breaking \cite{NGLV}, and some simplified vector field models called ``bumblebee" models have addressed the phenomenon at a more fundamental level \cite{BBEE}. Spontaneous Lorentz breaking is mandated by compatibility with general relativity in Riemannian spacetime, however explicit breaking can be reconciled if we expand to Riemann-Finsler geometries. The interpretation for this scenario would be to consider relativistic particles as following trajectories governed by a pseudo-Riemann-Finsler metric \cite{finsler}.
 
The SME is formulated as an effective field theory, independent of any underlying model, which allows testing of any type of Lorentz or CPT violation. It describes the effects of nonzero CPT-violating vacuum expectation values on conventional fields of the SM. The effects are represented with CPT-violating coupling coefficients of the SM fields to the nonzero background \cite{KyCol}. The violation of CPT also implies Lorentz violation \cite{greenberg}. These coefficients are testable at available energy scales, and bounds have been established in all sectors of the SM. These are published in data tables updated yearly \cite{tables}. 

The framework of Lorentz and CPT violation was incorporated into general relativity with \cite{KygnrelSME}, which also expanded experimental searches to short-range gravity \cite{shrgrav} and gravitational-waves \cite{grawa}. Some recent work uses bounds on Lorentz violation to constrain other possible subtle effects such as torsion \cite{torsion} and nonmetricity \cite{nonmet}.

Investigations here address the minimal SME which is power counting renormalizable, constraining the mass dimension of the SME coefficients. A nonminimal expansion of this to fermions with operators of arbitrary mass dimension was done in \cite{nonmin}.

In establishing SME phenomenology for neutral mesons, the traditional SM-based formalism has to be extended to include effects of the CPT-violating background. This formalism can be applied to design CPT tests consistent with the symmetry-breaking mechanism and the physics involved. 

Experiments with neutral mesons were the first to place bounds on an SME coefficient. It is the best experimental ground to study a particular quark-sector coefficient. Early SME investigations used the constant parameter defined in terms of the components of the effective hamiltonian describing the meson oscillations. This parameter describes CPT violation as mass and decay rate asymmetries between particle and antiparticle.  
 
However, in the proper adaptation of the SME phenomenology it was shown that if the CPT violation is rooted in coupling to a constant background, resulting from spontaneous symmetry breaking, the phenomenological parameter has to be momentum dependent to be consistent with quantum field theory. Based on this insight, a new formalism was developed in  \cite{NMPhKy01,NMPhKy99,NMPhKy98}. It addresses the issue that the laboratory moves compared to the constant background. It also takes into account momentum variations due to individual particle kinematics. The formalism is flavor dependent, since coupling to the background is specific to each quark field. 

The present work offers new ways to apply the existing neutral-meson SME phenomenology to the specific study of correlated meson decays. The analysis involves the definition of new asymmetries in the momentum-dependent approach to isolate the phenomenological parameter. Bounding the direction-dependent parameter places limits on the components of the SME coefficient, and provides new insights about the related physics.

The SME approach was used in experiments performed for over a decade in all neutral meson systems. The different collaborations contributed different constraints on various components, or combinations of components of the quark sector SME coefficient. As indicated above, the detectors fall into two categories: detectors involving uncorrelated mesons with high boost, and those using quantum correlated pairs produced nearly at rest. The latter are the meson factories, which have the advantage of low background and high precision. 

This distinction, however, is less clear now, with LHCb capable of limiting all four components of the SME coefficient with high precision, previously available only from meson factories \cite{LHCb16}. Similarly impressive improvements have been seen at D$\emptyset$ \cite{BestD015}. Meanwhile, KLOE has achieved comparable bounds to high-boost experiments, and placed component-by-component limits on the SME coefficient \cite{KLOE14}. It also did important measurements for quantum coherence of the correlated decays \cite{kaoncorrel}. 

The high-boost kaon experiments used to calculate bounds in Ref.\ \cite{NMPhKy98} and results from KTeV \cite{KTeV} still hold as the best limits on SME coefficients, due to their boost factor. For the $B$ factories only BaBar set SME limits, collecting data between 1999 and 2008 \cite{Babar08} while Belle ran its tests between 1998 and 2010 in the context of traditional searches. Latest bounds from BaBar come from evaluating existing data \cite{schubert}. There are separate studies of Lorentz violation involving the top quark \cite{Berger}. In general, experiments in the last few years have achieved bounds of 1 to 3 orders of magnitude better than before for the components of the CPT-violating coefficient. 

This paper focuses on the updated Belle detector. Earlier SME formalism is expanded to give explicit results for studying CPT in meson factories. Currently, Belle II provides the only ongoing data collection on $B$ mesons produced in an entangled state. Using the findings presented here, experiments can probe SME-type CPT violation in all its facets. It is shown that all four components of the relevant quark sector coefficient can be bound. Due to better detection methods and higher luminosity, improved bounds are expected. Since the relevant SME coefficient depends on quark flavor \cite{KyPotting}, and possibly on mass or lepton, baryon number \cite{shade}, the $B$ factory searches could be significant.  
 
The approach here discusses a basic difference between original phenomenology assuming only mass and decay rate differences of particle and antiparticle, and the framework also considering directional dependence. The former is signified by a constant phenomenological parameter. The analysis throughout the paper shows that by focusing on correlated decay analysis specific to $B$ mesons, it is possible to isolate the features of momentum-dependence at the detection level.

This difference manifests in $B$ factories in a unique way. Some of the general correlated decay study, however, is equally applicable for experiments at KLOE. Investigating it is of particular importance both, experimentally and theoretically. 

For full understanding, the time development before the decay of the first $B$ meson is properly included, leading to modified decay rate asymmetries. Since the time of the first decay cannot be measured accurately, detection has to rely on an analysis of kinematics, quantum correlations and decay rate asymmetries.  

One signal of a direction-dependent effect can be a nonzero asymmetry between conjugate same-flavor decay rates, which vanishes to first order in a formalism with constant CPT violating parameter. Due to the characteristics of the kinematics of $B$ meson propagation, it can also be observed with detector binning of flavor-specific decays products. 

A third test is a study of possible decoherence effects. It is shown here that propagation considered from the production of the $B$ meson pair in a nonzero background can produce disentangled states to second-order in the CPT violating parameter. These investigations provide a more fundamental test of the underlying physics.   

The paper starts in Sec.\ \ref{momdepphen} with a review of the formalism developed previously for momentum dependent CPT violation searches. This section closely follows the now-standard SME phenomenology for neutral mesons, summarizing key equations and the connection to the relevant SME coefficient as introduced by Refs.\ \cite{NMPhKy01,NMPhKy99,KyPotting}. Section\ \ref{Genbas} gives the minimal SME lagrangian with the testable term, the hamiltonian parametrization, eigenstates and basic time development of the meson states and connects to it the SME coefficients. Section \ \ref{Mesfac} gives an overview of the relevant entangled state and amplitude for correlated decays. Following it Sec.\ \ref{newsmesfac} presents specific adaptations of the phenomenology to Belle II type searches. The reader familiar with the neutral meson experiments can advance to this section. Section\ \ref{Comphen} discusses formal and fundamental differences in phenomenology of BaBar, Belle, and the SME, summarized in Table \ref{table1}. In Sec.\ \ref{SemlepDecrat} the specifics of semileptonic decays in the approximation of small CPT violating parameter are discussed. Section \ \ref{dirdep} addresses issues of geography and kinematics, and Sec.\ \ref{decoh} analyzes possible decoherence effects. Section \ \ref{sum} gives an outlook for possible investigations for the new Belle II detector. 
\section{Momentum-dependent neutral-meson phenomenology}
\label{momdepphen}
\subsection{General basics}
\label{Genbas}
In this section the neutral-meson phenomenology is presented briefly in the SME framework. It follows the original formalism created specifically for SME-based neutral-meson searches. It is a short but complete summary applied later in the paper to searches in a meson factory. The reader is referred for a detailed discussion to \cite{KyCol,NMPhKy98,NMPhKy99,NMPhKy01}. 
  
To start, a simplified form of the lagrangian of the minimal SME is given, containing only renormalizable terms. For massive spin-$\frac{1}{2}$ fermions it has the general form
\beq
\cl = \half i\overline{\psi} \Ga^\nu
\stackrel{\leftrightarrow}
{\prt}_{\nu}
\hspace{-.1cm}{\psi}
-\overline{\psi}M{\psi},
\label{lagr}
\eeq
where the extension defines $\Ga$ and $M$ including the SME coefficients as
\bea
{\Ga}^{\nu}&:=&{\ga}^{\nu}+c^{\mu \nu}
{\ga}_{\mu}+d^{\mu \nu}{\ga}_{5} {\ga}_{\mu}
\nonumber\\
&&\qquad
+e^{\nu}+if^{\nu}{\ga}_{5}
+\frac{1}{2}g^{\la \mu \nu}
{\sigma}_{\la \mu},
\label{Gamma}
\eea
and
\beq
M:=m+a_{\mu}{\ga}^{\mu}+b_{\mu}{\ga}_{5}
{\ga}^{\mu}+\frac{1}{2}H^{\mu \nu}
{\sigma}_{\mu \nu}.
\label{Mass}
\eeq
Here, $\ga_5$, $\ga^{\mu}$,$\ga_5\ga^{\mu}$, $\sigma^{\mu \nu}$ represent conventional gamma matrices, while 
$a_{\mu}$, $b_{\mu}$, $c_{\mu\nu}$, $\ldots$, $H_{\mu \nu}$ are determined by background expectation values of Lorentz tensors 
arising from the spontaneous Lorentz breaking \cite{KyCol}. Comparing properties of these terms to the properties of the neutral-meson oscillations a particular coefficient of the SME was identified as one that can be tightly constrained only in this system \cite{NMPhKy99,NMPhKy98}.

The neutral-meson pairs differ only in flavor and present a unique physical phenomenon where particle and antiparticle can oscillate into each other via weak processes, providing a sensitive interferometric probe for CPT tests. The standard approach for describing the coupled oscillation of the mesons uses a Schr\"odinger-type equation for a linear combination of the wave functions comprised of $B^0$ and $\overline{B^0}$, represented as a two-component object $\Ps(t)$. Here $B^0$ can represent $B_{d}^0$ or $B_{s}^0$ but it can stand for any neutral meson. 
The time evolution is described by a 2$\times$2 effective hamiltonian $\La$,  
\beq
i\prt_t \Ps = \La \Ps ,
\label{schreq}
\eeq
for a full treatment of the quantum system see for instance Ref. \cite{BigiSanda}. There are also classical models illustrating the physics \cite{classmod}.

The hamiltonian \rf{schreq} has many different parametrizations. In Ref.\ \cite{NMPhKy01} a comparative summary is given and a convenient parametrization is presented for the SME searches. Following that parametrization $\La$ has the form
\beq
\La = 
\half \De\la
\left( \begin{array}{lr}
U + \xi 
& 
\quad VW^{-1} 
\\ & \\
VW \quad 
& 
U - \xi 
\end{array}
\right).
\label{uvwx}
\eeq
$U,V,W,\xi$ are all complex parameters with $W = w \exp (i\om)$, while $\xi= \Re\xi + i\Im\xi$. This allows description of spacetime symmetry violations in this system with four independent dimensionless real phenomenological parameters, which are independent of phase conventions and of the particular model. 

In the hamiltonian \rf{uvwx}, off-diagonal components of $\La$ control the flavor oscillations between $B^0$ and $\overline{B^0}$ and are described by the CP violation parameter $w$. In case of T symmetry, $w = 1$.

Meanwhile, indirect CPT violation occurs if and only if the difference of the diagonal elements in the effective hamiltonian above is nonzero; $\De\La = \La_{11}-\La_{22}\neq 0$. This property compels us to look for flavor diagonal SME terms to be tested. Inspecting $\La$, it is clear that CPT violation is described by a nonzero $\xi$.
The final step is to express the quantities $w$, $\xi$ and $V$ in terms of the components of the hamiltonian $\La$, giving
\beq
w = \sqrt{|\La_{21}/\La_{12}|},
\quad
\xi = \De\La/\De\la ,
\quad
V = \sqrt{1-\xi^{2}}.
\label{wVcalc}
\eeq

The physical propagating states of the neutral $B$ system, $\ket{B_L}$ and $\ket{B_H}$, are the eigenstates of $\La$. The  neutral $B$ particles are produced as strong interaction eigenstates with definite parity, carrying the same parity eigenvalue. Hence they remain parity eigenstates. The corresponding SME coefficient has to match this behavior and has to be parity preserving while violating charge conjugation.

Connecting to the general SME lagrangian a single term was found that is flavor diagonal and is parity preserving while violating C symmetry. The relevant term is the flavor dependent $- a^q_0 \overline{q} \ga^0 q$, where $a^q_0$ is the zeroth component of a vector coefficient in the quark sector describing coupling to the background, stemming from the spontaneous Lorentz symmetry breaking \cite{NMPhKy98,NMPhKy01}.

In Minkowski spacetime the coefficient is $a^q_0$, in itself, is undetectable with a single flavor fermion and could be arbitrarily large. It could be observed in the presence of gravity, but there it is countershaded by the weakness of the gravitational coupling \cite{shade}. Instead the difference $\De a_0 \equiv a^{q_1}_0 - a^{q_2}_0$ is observed, which can only be done in flavor changing of neutral meson or neutrino oscillations. In neutral meson, as well as neutrino oscillations an interferometric effect allows placing tight bounds on $\De a_0$. Anomalous oscillations have been reported for both systems \cite{anomne},\cite{anomme}.

Neutrino oscillations are more involved than meson oscillations. This area constitutes its own field in SME searches. Theoretical descriptions are found in \cite{nebasic}, for the nonminimal extension \cite{nebasicnonmin}. 

Bounding Lorentz violation in the neutrino sectors can be done by finding limits of sidereal variations of oscillations or distortions of the oscillation spectra. A large number of operators are needed to describe Lorentz violation. Some of these are probed looking at sidereal variations in neutrino oscillations, such as astrophysical \cite{neastro}, reactor \cite{nereact}, accelerator based short-baseline \cite{neshort} and long-baseline \cite{nelong} investigations, atmospheric \cite{neatm} and solar neutrino oscillations \cite{nesolar}. Countershaded violations involving oscillation-free operators are investigated in $\beta$ and double $\beta$ decay \cite{nebeta}.

In neutral meson oscillations, the difference of the diagonal elements is proportional to $\De a_0$ for the valence quarks of different flavor, and can be detected with the proper decay rate asymmetry. The difference is the consequence of an energy shift of the rest energies due to coupling to the CPT violating background, and is equal but opposite for the meson and its anti-meson. Its manifestation is a corresponding difference in lifetime and/or mass of particle and anti-particle. This translates into a relationship between the SME coefficients and the difference in diagonal elements $\La$ given by  $\De\La \approx \De a_0 \equiv a^{q_1}_0 - a^{q_2}_0$, for the two valence quarks in the $B^0$ meson, with ${q_1}$ and ${q_2}$ indicating the flavors \cite{KyPotting}. 

In traditional phenomenology this parameter is defined for particles at rest where $\De a_0$ is rotationally invariant. However, for boosted mesons the formalism changes and momentum dependence has to be considered as shown in detail in Sec.\ \ref{dirdep}. In general
\beq
\De\La \approx \be^\mu \De a_\mu ,
\label{delambda}
\eeq
where $\be^\mu = \ga (1, \vec \be )$ is the four-velocity of the meson state, defined as usual by $\vec{v}=\frac{d\vec{x}}{dt}$. $\De a_\mu$ is the symmetry-violating coefficient given in the observer frame. 

This dependence on boost and angular distribution is a key experimental issue in the SME phenomenology. While high boost experiments deliver enhanced signal for CPT violation, here the focus is on directional dependence. The full treatment of this dependence is postponed for Sec.\ \ref{dirdep}. The only fact accented here is that the correlated wave function, decay amplitudes, decay rate probabilities and asymmetries carry dependence on direction and the sidereal time of Earth's rotation relative to the constant background. The time development of the states is modified accordingly.

Comparing this to expressions \rf{wVcalc}, the connection between the phenomenological parameter $\xi$ and $\De a_\mu$ is established in the form
\beq
\xi = \De\La/\De\la \approx \fr{\be^\mu \De a_\mu}{\De\la},
\label{xilambda}
\eeq 
where $\be^\mu$ denotes the four-velocity. 

For a full study of the oscillation and propagation the time development of the eigenstates of hamiltonian \rf{schreq} are needed. They evolve in time according to
\bea
\ket{B_L(t)}&=&\exp (-i\la_Lt) \ket{B_L},
\nonumber\\
\ket{B_H(t)}&=&\exp (-i\la_Ht) \ket{B_H},
\label{timevol}
\eea
where the eigenvalues are composed of the propagating masses $m_L$, $m_H$ and decay  rates $\Ga_L$, $\Ga_H$ according to 
\beq
\la_L \equiv m_L - \half i \ga_L , \quad 
\la_H \equiv m_H - \half i \ga_H .
\label{mga}
\eeq

One can also define the sum and difference of the eigenvalues 
\bea
\la &\equiv &\la_L + \la_H = m - \half i \ga ,
\nonumber\\
\De \la &\equiv &\la_L - \la_H = - \De m - \half i \De \ga ,
\label{lambdas}
\eea
where for the masses and decay rates we define 
\bea 
m = m_L + m_H,
\quad 
\De m = m_H - m_L,
\nonumber\\
\ga = \ga_H + \ga_L ,
\quad
\De \ga = \ga_H - \ga_L .
\label{massdec}
\eea
These  quantities play a role in the oscillation characteristics of the mesons and without spacetime-symmetry violations would entirely determine the time development. Their values are also important for experiments, since they characterize the interference and the difference in accessibility to discrete symmetry measurements for neutral mesons of different quark content. They also greatly influence design requirements for their respective detectors. 

Finally, the normalized physical states in the above formalism can be expressed in terms of the strong eigenstates. This allows to continue on to the full time dependence of the correlated neutral meson states in the upcoming sections. The relation according to the hamiltonian is
\bea
\ket{B_L} &\sim & \ket{B^0} + (1 - \xi) W/V \ket{\overline{B^0}} , 
\nonumber \\
\ket{B_H} &\sim & \ket{B^0} -(1 + \xi) W/V \ket{\overline{B^0}} .
\label{statesdef}
\eea

Once the full time dependence is known, decay rate probabilities are determined for the coherent states. Their asymmetries place limits on the phenomenological parameter. Those limits in turn give constraints on the SME coefficient.  
\subsection{Meson factory basics}
\label{Mesfac}
This section presents the phenomenology of correlated meson decays characteristic for meson factories. 
The first step is to determine the time-dependent states  $B^{0}(t)$, $\overline{B^{0}}(t)$ as
\bea
\bra {B^0(t, \hat t, \vec p)} &=& (C + S \xi)\bra {B^{0}} + (SVW)\bra {\overline{B^0}}, 
\nonumber\\
\bra {\overline{B^0}(t, \hat t, \vec p)} &=& (SVW^{-1})\bra {B^{0}} + (C - S \xi)\bra {\overline{B^0}},
\eea
where functions $C$ and $S$ of the meson proper time $t$ are defined as
\bea
C &=&
\cos (\half \De\la t)
\exp(-\half i \la t)
\nonumber\\
&=&
\half(e^{-i\la_Lt} + e^{-i\la_Ht}),
\nonumber\\
S &=&
- i \sin (\half \De\la t)
\exp(-\half i \la t)
\nonumber\\
&=&
\half(e^{-i\la_Lt} - e^{-i\la_Ht}).
\label{CS}
\eea

Later in this section, a detailed correspondence is given to relate the formalism here to that of BaBar and Belle. However, to fully understand these relations it is better to first write down the correlated decay amplitude, shown below. After the production, the neutral $B$ meson pair emerges in a coherent state of quantum entanglement. This correlation can be described by
\beq
\ket{i} = 
\fr 1{\sqrt{2}}
\bigl(
\ket{B^0(+)} \ket{\overline{B^0}(-)}
-\ket{B^0(-)} \ket{\overline{B^0}(+)}
\bigr),
\label{correlst}
\eeq
where $(+)$ and $(-)$ refer to the opposing directions in the quarkonium rest frame in which the particle pair is moving. 

Following Ref.\ \cite{NMPhKy01}, the amplitudes $F_a$ and $\overline F_{a}$, $a = 1,2$ into final states $f_1$ at time $t_1$, and into $f_2$ at time $t_2$ are given by 
\beq
\bra{f_a}T\ket{B^0}
= F_a, 
\qquad 
\bra{f_a}T\ket{\overline{B^0}}
= \overline F_{a}. 
\label{genamps}
\eeq
Time  $t_2$ is assumed to be the time of the second decay. The dependence on both decay times is observed throughout, to account for possible physics stemming from interaction with the background before the first decay. 

As discussed above, in the approach used in this work, these amplitudes are a function of three-momentum $\vec p_1$ and $\vec p_2$ of the two particles. The four-momentum is an eigenvalue of the translation operator and is conserved. As indicated in Eq.\ \rf{delambda} above, the difference in the diagonal elements of the effective hamiltonian depends on the four-velocity. Note that for the case of directionally-dependent Lorentz violation the relation of the velocity and momentum generally differs from that in the special relativistic case. This modification by the background is only second order in any of the Lorentz-violating coefficients and hence here it is neglected. For an explicit relationship between velocity and the canonical momentum in the SME framework the reader is referred to \cite{KyCol}. The relevant details will be discussed in Sec.\ \ref{dirdep}.  

Changes in orientation relative to the constant background resulting from the Earth's daily rotation is included as sidereal time dependence of the amplitude, denoted here as $\hat t$. While the sidereal time is considered fixed during the decay process, variations with sidereal time overall provide the means of constraining some of the individual components of the SME coefficient.

With these considerations the probability amplitude $A_{f_1f_2}$ for the decays becomes
\bea
& A_{f_1f_2} &
\equiv 
A_{f_1f_2}(t_1,t_2,\hat t,\vec p_1,\vec p_2)
=
\bra{f_1 f_2}T\ket{i} 
\nonumber\\
&=&
\fr 1{\sqrt{2}}
\bigl[
\bra{f_1}T\ket{B^0(t_1,\hat t,\vec p_1)}
\bra{f_2}T\ket{\overline{B^0}(t_2,\hat t,\vec p_2)}
\nonumber\\
&&
- \bra{f_1}T\ket{\overline{B^0}(t_1,\hat t,\vec p_1)}
\bra{f_2}T\ket{B^0(t_2,\hat t,\vec p_2)}
\bigr].
\label{amp12}
\eea 
In the SME formalism with definitions \rf{genamps}, the amplitude can be expanded
using the separate time development functions $C_a = C(t_a)$, $S_a = S(t_a)$, $a=1,2$. This gives the expression for the amplitude in the form
\bea
&A_{f_1f_2}& = 
\fr 1{\sqrt{2}}
\bigl[
F_1\ol F_2(\xi_1S_1C_2 - \xi_2S_2C_1 + C_1C_2 
\nonumber\\
&&
\hbox{\hskip-20pt}
\hspace{2 cm} - (\xi_1\xi_2+V_1V_2)S_1S_2)
\nonumber\\
&&
\hbox{\hskip-20pt}
+ F_2\ol F_1
(\xi_1S_1C_2 - \xi_2S_2C_1 - C_1C_2
\nonumber\\
&&
\hbox{\hskip-20pt}
\hspace{2 cm} + (\xi_1\xi_2+V_1V_2)S_1S_2)
\nonumber\\
&&
\hbox{\hskip-20pt}
+ F_1 F_2 W^{-1}(V_2 C_1 S_2 - V_1 S_1 C_2 
\nonumber\\
&&
\hbox{\hskip-20pt}
\hspace{2 cm} + (\xi_1 V_2 - \xi_2 V_1) S_1 S_2)
\nonumber\\
&&
\hbox{\hskip-20pt}
+ \ol F_1\ol F_2 W(V_1 S_1 C_2 - V_2 C_1 S_2
\nonumber\\
&&
\hbox{\hskip-20pt}
\hspace{2 cm} + (\xi_1 V_2 - \xi_2 V_1) S_1 S_2)
\bigr].
\label{expampl12}
\eea
This general form implicitly carries all information about the possible decays. The quantities $V_1\equiv\sqrt{1-\xi_{1}^{2}}$, $V_2\equiv\sqrt{1-\xi_{2}^{2}}$ are defined in terms of $\xi_1(\vec p_1)$, $\xi_2(\vec p_2)$, while $W= w\exp(i\om)$ as in \rf{wVcalc}. 
\section{New searches in meson factories}
\label{newsmesfac}
\subsection{Comparison of phenomenologies}
\label{Comphen}
Expression \rf{expampl12} shows an extended formalism, different from those found in the BaBar and Belle papers, which are based on a phenomenology developed for the particular experimental methods adapted in $B$ factories and is detailed in, for example, Ref. \cite{Babar04}.\ Connecting these formalisms requires two things to be taken into consideration. 

The first is the correspondence in notation, which is simply a matter of denoting certain functions and quantities with different symbols. Such notational difference can be seen for instance in the decay amplitudes for the decay modes, defined for BaBar and Belle as $A_{1,2}$ or by $A_{tag},A_{rec}$ corresponding to $F_{1,2}$ here. The functions $C$ and $S$ correspond to functions $g_{+,-}$ in BaBar's notation. The correlated double decay amplitude in BaBar's formalism is divided into symmetry violating and non-violating by introducing $a_{+,-}$. They are combined with the respective time functions $g_{+}$,$g_{-}$ for the time development. The terms $a_{+,-}$ are denoted at Belle as $\eta_{+,-}$. Important is also that the CP violating parameter $W$ corresponds to $\frac{q}{p}$ and the CPT violating parameter $\xi$ to $-z$.
\begin{table}
\begin{center}
\tabcolsep8pt
\begin{tabular}{|c|c|}
\hline
 SME & BaBar, Belle II\\
\hline
\hline
$\xi_{1} = \xi_{1}(t_{1},p_{1})$ & 0 \\
\hline
$\xi_{2} = \xi_{2}(t_{2},p_{2})$ & $-z$ \\
\hline
$V_{1} \equiv \sqrt{1-\xi_{1}^{2}}$ & 1 \\
\hline
$V_{2} \equiv \sqrt{1-\xi_{2}^{2}}$ & $\sqrt{1-z^{2}}$ \\
\hline
$W = w e^{i\omega}$ & $W = \frac{q}{p}$ \\
\hline
$t_{1}$ time of first decay & $t_{0} = 0$ \\
\hline
$t_{2}$ time of second decay &	$\Delta t = t_{2} - t_{1}$ \\
\hline
$C_{1} = \half(e^{-i\la_Lt} + e^{-i\la_Ht}), t \leq t_{1}$  & time development before decay \\
\hline
$S_{1} = \half(e^{-i\la_Lt} - e^{-i\la_Ht}), t \leq t_{1}$  & time development before decay \\
\hline
$C(\Delta t) = \half(e^{-i\la_L\Delta t} + e^{-i\la_H\Delta t})$  &	$g_{+} = \half(e^{-i\omega_Lt} + e^{-i\omega_Ht})$ \\
\hline
$S(\Delta t) = \half(e^{-i\la_L\Delta t} - e^{-i\la_H\Delta t})$  &	$g_{-} = \half(e^{-i\omega_Lt} - e^{-i\omega_Ht})$ \\
\hline
$\la_{L,H}$ &	$\omega_{L,H}$ \\
\hline
$F_{1,2}$  &  $A_{1,2}$	 \\
\hline
$\overline{F}_{1,2}$ & $\overline{A}_{1,2}$	 \\
\hline
\end{tabular}
\vskip -5pt
\caption{\label{table1}
Comparison of phenomenological descriptions.}
\end{center}
\vskip -10pt
\end{table}

The second is more fundamental, and is based in the underlying physics due to the different mechanism by which violation of CPT symmetry can occur as compared to CP violation. At the $B$ factories, the time for the decay of the first meson is difficult to measure, and with flavor tagging and using only the difference of decay times, it is unnecessary to determine. The oscillations up to the first decay point are correlated and once the flavor of one $B$ meson is known the other can be inferred as well.

Searches are done by taking one of the mesons to be the tagging meson and fully reconstructing the other. The zero of the decay time is considered to be at the time of the first decay, and further time evolution can be taken as a function of the time measured after the first decay until the time of the second one. This can be experimentally determined and is denoted by $\Delta t$ (sometimes just $t$). This is a great simplification for the formalism.

Meanwhile, in case of CPT violation, the background can have an effect on the meson-antimeson pair depending on their momenta immediately after production. Due to this, the time development in the SME formalism has to be described from the moment the particles are produced in the decay of $\Upsilon(4S)$, analogous to when quantum correlations are studied \cite{BelleEPR07}. Here time variables are defined as $\Delta t=t_{2}-t_{1}$ and $t=t_{2}+t_{1}$. $C(t)$ and $S(t)$ have separate time dependence for the two mesons marked as $C_1 = C(t_1)$, $C_2 = C(t_2)$, $S_1 = S(t_1)$, $S_2 = S(t_2)$. The CPT violation parameter also carries separate indeces denoted as $\xi_1(\hat t, \vec p_{1})$ and $\xi_2(\hat t, \vec p_{2})$ to account for the momentum dependence. A summary of all correspondences is given in Table \ref{table1}. It should be noted that that neutral meson searches have been done with a number of different parametrizations. A summary of them is given in  \cite{NMPhKy01} for the most generally used notations or found in relevant works giving relation to the BaBar notation as, for instance, in \cite{Kundu1}.  

For an explicit illustration on how to use Table \ref{table1} an example is given for the case of no CPT violation below. In that case $\xi_{1}=\xi_{2}=0$ and $V_{1}=V_{2}=1$, while the amplitude and CP violating parameter correspondences are explicitly shown connecting to the formalism at $B$ factories as
\bea
&A_{1}& \rightarrow F_{1} ,\quad  A_{2} \rightarrow F_{2},  
\nonumber\\
&&
\hbox{\hskip-20pt} 
\vspace{0.3 cm}
\overline{A}_{1} \equiv \overline{F}_{1} ,\quad  \overline{A}_{2} \equiv \overline{F}_{2},
\nonumber\\
&&
\hbox{\hskip-20pt}
\vspace{0.3 cm}
\frac{p}{q} \rightarrow W^{-1},\quad \frac{q}{p} \rightarrow W.
\label{corresp}
\eea
The CPT preserving amplitude expressed in the SME formalism is
\bea
A_{f_1f_2} &=& 
\fr 1{\sqrt{2}}
\bigl[
(F_1\ol F_2)C - (F_2\ol F_1)C 
\nonumber\\
&&
\hbox{\hskip-20pt}
+ (F_1 F_2 W^{-1})S + (\ol F_1\ol F_2 W)S
\bigr],
\label{smeamp}
\eea
while the same amplitude expressed in BaBar's notation has the form
\bea
A_{f_1f_2} &=& 
\fr 1{\sqrt{2}}
\bigl[
(A_1\ol A_2)g_{+} - (A_2\ol A_1)g_{+} 
\nonumber\\
&&
\hbox{\hskip-20pt}
+ (A_1 A_2 \frac{p}{q})g_{-} + (\ol A_1\ol A_2 \frac{q}{p})g_{-}
\bigr].
\label{babamp}
\eea
\subsection{Semileptonic decay rates}
\label{SemlepDecrat}
The general amplitude of Eq.\ \rf{expampl12} is valid for any size CPT violations. Expanding it with the specific time dependent oscillation and decay functions, however, is involved. Here it is only done in a small $\xi$ approximation to first order in $\xi$. As we will see, in this approximation decoherence effects are not a concern, but they do appear in an extension to second order. The following definitions are used for clarity,
\beq
\xi = \xi_1 + \xi_2,
\quad
\De \xi = \xi_1 - \xi_2.
\label{xidexi}
\eeq
For the overall time evolution factor $k(t)\equiv \exp(-i \cdot \lambda t/2)$ is defined.

This gives the amplitude for final states $f_1$ and $f_2$
\bea
&A_{f_1f_2}& = 
\fr 1{\sqrt{2}} k(t)
\lbrace
\nonumber\\
&&
\hbox{\hskip-20pt}
F_1\ol F_2 \bigl[\cos \frac{\Delta\lambda\Delta t}{2} +
\nonumber\\
&&
\hbox{\hskip-20pt}
\hspace{2 cm}\frac{i}{2}(\xi \sin\frac{\Delta\lambda\Delta t}{2} - \Delta\xi \sin\frac{\Delta\lambda t}{2})\bigr]
\vspace{0.4 cm}
\nonumber\\
&&
\hbox{\hskip-20pt}
+ F_2\ol F_1 \bigl[- \cos \frac{\Delta\lambda\Delta t}{2}
\nonumber\\
&&
\hbox{\hskip-20pt} 
\hspace{2 cm} + \frac{i}{2}(\xi \sin\frac{\Delta\lambda\Delta t}{2} - \Delta\xi \sin\frac{\Delta\lambda t}{2})\bigr]
\vspace{0.4 cm}
\nonumber\\
&&
\hbox{\hskip-20pt}
+ F_1 F_2 W^{-1}\bigl[-i \sin\frac{\Delta\lambda\Delta t}{2}
\nonumber\\
&&
\hbox{\hskip-20pt}
\hspace{2 cm} + \frac{1}{2}\Delta\xi\left(\cos\frac{\Delta\lambda t}{2} - \cos \frac{\Delta\lambda\Delta t}{2} \right)\bigr]
\vspace{0.4 cm}
\nonumber\\
&&
\hbox{\hskip-20pt}
+ \ol F_1\ol F_2 W
\bigl[i \sin\frac{\Delta\lambda\Delta t}{2}
\nonumber\\
&&
\hbox{\hskip-20pt}
\hspace{2 cm}+ \frac{1}{2}\Delta\xi\left(\cos\frac{\Delta\lambda t}{2} - \cos \frac{\Delta\lambda\Delta t}{2} \right)\bigr]
\rbrace.
\nonumber\\
&&
\hbox{\hskip-20pt}
\label{amplambda}
\eea
Here, the detailed oscillation and decay information is still contained in $\lambda$ and $\Delta\lambda$. This amplitude will be calculated more explicitly later, with the simplification of focusing only on semileptonic decays. At this stage, however, some important points can be made. 

The first two terms pertain to conjugate decay products occurring in different time order. Their first time development function  depends only on decay time differences $\Delta t$ and indicates physics without symmetry violations. This term switches sign as the decay products switch and is even for $\Delta t$.
 
In the parentheses the two other terms describe time evolution combined with the CPT violating parameter. They are both odd functions of time but the term with $\xi$ depends on the traditional time variable $\Delta t = t_{2} - t_{1}$, while the term containing $\Delta\xi$ is a function of $t = t_{2} + t_{1}$. This term contains the time before the first decay and cannot be absorbed in the usual formalism based on measuring only $\Delta t$. The parameter $\Delta\xi$ is tied to momentum dependence and is only present in the SME phenomenology. Since these terms represent opposite flavor decays, a nonzero amplitude for the states to coincide is consistent with quantum entanglement.

All terms, where time development before the first decay cannot be removed using $\Delta t$, are connected to only $\Delta\xi$ in any case. It stems from the fact that particles with different momenta interact different with the symmetry-breaking background. These terms do not vanish as $\Delta t$ goes to zero. As will be discussed in Sec. \ref{decoh}, they carry special relevance for the last two parts of amplitude \rf{amplambda}, expressing decays into same-flavor states. Those terms are expected to be zero for $\Delta t=0$ for same-flavor decays occurring at the same time. 

The decay amplitudes into the same-flavor modes have a first term that is odd in $\Delta t$ and is imaginary. It describes oscillation without CPT violation. The next two terms only contain $\Delta\xi$, indicating, that to first order in $\xi$, the same-flavor symmetry violation would be zero. Its appearance in correlated decays is a special feature of the directionally dependent phenomenology. By inspection ones sees, that to first order in $\Delta\xi$, the decay rate probability is zero for the two particles decaying into the same mode at the same time, because the multiplying sine function is zero at $\Delta t=0$. That means that at this level of approximation no decoherence occurs.
  
Turning attention to semileptonic decays further analysis can be done. Note that decays to CP eigenstates are discussed in detail in searches done at CERN \cite{Tilburgsum}. There are other broader decay mode specific analyses available for intance by \cite{Kundu1,Kundu23}.  For flavor specific semileptonic decays the basic transition amplitudes can be rewritten as
\bea
\bra{f}T\ket{P^0}
= F , &
\qquad 
\bra{f}T\ket{\overline{P^0}}
= 0,
\nonumber \\
\bra {\overline f}T\ket {\overline{P^0}}
= \overline F, &
\qquad 
\bra {\overline f}T\ket{P^0}
= 0.
\label{semlepF}
\eea

The time development functions are somewhat lengthy and obscure their actual effect on the amplitudes. So for further clarity functions $h_{1}(t) \equiv \sin \De \la t/2$ and $h_{2}(t) \equiv \cos \De \la t/2$ are also defined. This gives the following form in the momentum dependent formalism for the semileptonic decay amplitudes, 
\bea
A_{f \ol f} &=&  
\frac{(F \ol F)}{2\sqrt{2}}
{k(t)}\left[i\xi h_{1}(\Delta t) - i\Delta\xi h_{1}(t) + 2 h_{2}(\Delta t)\right],
\nonumber\\
\hbox{\hskip-20pt}
A_{\ol f f} &=&  
\frac{(\ol F F)}{2\sqrt{2}}
{k(t)}\left[i\xi h_{1}(\Delta t) - i\Delta\xi h_{1}(t) - 2 h_{2}(\Delta t)\right],
\nonumber\\
\hbox{\hskip-20pt}
A_{f f} &=&
\frac{F^{2} W^{-1}}{2\sqrt{2}}
{k(t)}\left[\Delta\xi h_{2}(t)-\Delta\xi h_{2}(\Delta t) - 2 i h_{1}(\Delta t)\right],
\nonumber\\
\hbox{\hskip-20pt}
A_{\ol f \ol f} &=&
\frac{\ol F ^{2} W}{2\sqrt{2}}
{k(t)}\left[\Delta\xi h_{2}(t)-\Delta\xi h_{2}(\Delta t) + 2 i h_{1}(\Delta t)\right].
\nonumber\\
&&
\hbox{\hskip-20pt}
\label{semlepamp}
\eea
The decay rates are again calculated only to first order in $\xi$ and $\Delta \xi$. 

Note that these explicit forms now show the dependence on the imaginary and real part of $\xi$. In the SME framework there is a constraint between them founded in the fact that the perturbation hamiltonian is hermitian, so $\De\La$ is real.
\beq
\Re \xi = - 2 \De m \Im \xi/\De \ga.
\label{reim}
\eeq
Relationship \rf{reim} is not used in general here. However, it can be helpful in a treatment that assumes $\Delta\gamma$ to be zero and contains limits only on the value of  $\Im \xi$ and $\Im \Delta\xi$.

The flavor specific decay rate probabilities can now be calculated and take the form
\bea
P_{f \ol f} &=&   
k_{f \ol f}\bigl[\cosh \frac{\Delta\gamma\Delta t}{2} + \cos \Delta m \Delta t
\nonumber\\
&&
\hbox{\hskip-20pt}
+ \Im \xi \sin \Delta m \Delta t + \Re \xi \sinh\frac{\Delta\gamma \Delta t}{2}
\nonumber\\
&&
\hbox{\hskip-20pt}
- 2 \Im(\Delta \xi ^{\star} h^{\star}_{1}(t) h_{2}(\Delta t))\bigr],
\nonumber\\
&&
\hbox{\hskip-20pt}
\vspace{0.4 cm}
P_{\ol f f} = P_{f \ol f} (\Delta \xi \rightarrow - \Delta \xi, \xi \rightarrow - \xi),
\eea
\bea
P_{f f} &=&
k_{f f} \bigl[\cosh \frac{\Delta\gamma\Delta t}{2} - \cos \Delta m \Delta t
\nonumber\\
&&
\hbox{\hskip-20pt}
- \Im \Delta \xi \sin \Delta m \Delta t + \Re \Delta  \xi \sinh\frac{\Delta\gamma \Delta t}{2}
\nonumber\\
&&
\hbox{\hskip-20pt}
- 2 \Im(\Delta \xi ^{\star} h^{\star}_{2}(t) h_{1}(\Delta t))\bigr],
\nonumber\\
&&
\hbox{\hskip-20pt}
\vspace{0.4 cm} 
P_{\ol f \ol f} = P_{f f} (\Delta \xi \rightarrow - \Delta \xi, k_{f f} \rightarrow k_{\ol f \ol f}),
\label{semlepprob}
\eea
where $k_{\ol f f}, k_{f \ol f}, k_{f f}, k_{\ol f \ol f}$ are defined as:
\bea
k_{f \ol f}\equiv k_{\ol f f} &=& \frac{1}{4}(F \ol F)^{2}e^{\frac{-\gamma t}{2}},
\nonumber\\
&&
\hbox{\hskip-20pt}
\hspace{-2 cm}
k_{f f} = \frac{1}{4}(|F^{2}|)^{2}e^{\frac{-\gamma t}{2}}, \qquad k_{\ol f \ol f} = \frac{1}{4}(|\ol F^{2}|)^{2}e^{\frac{-\gamma t}{2}}.
\label{kff}
\eea
The last term in these expressions is complicated by the fact that it carries the time dependence before the first decay $(t_{1})$, along with functions depending on $\Delta t$. Its expansion is lengthy and the information contained is difficult to apply to any specific experimental scenario. 

To be able to make some important points, the functions $f_{i}=f_{i}(t_{1}, \Delta t), i=1...8$ are defined below. This allows an analysis of their properties and the influence on the probabilities. Some of those relevant properties of $f_{i}$ are summarized in Table \ref{table2}.
\bea
f_{1}&=&\sinh \frac{\Delta\gamma\Delta t}{2} \cos \Delta m t_{1} \cosh \frac{\Delta\gamma t_{1}}{2},
\nonumber\\
f_{2}&=&\sin \Delta m \Delta t \sin \Delta m t_{1} \sinh \frac{\Delta\gamma t_{1}}{2},
\nonumber\\
f_{3}&=&\cosh \frac{\Delta\gamma\Delta t}{2} \cos \Delta m t_{1} \sinh \frac{\Delta\gamma t_{1}}{2},
\nonumber\\
f_{4}&=&\cos \Delta m \Delta t \cos \Delta m t_{1} \sinh \frac{\Delta\gamma t_{1}}{2},
\nonumber\\
f_{5}&=&\sinh \frac{\Delta\gamma\Delta t}{2} \sin \Delta m t_{1} \sinh \frac{\Delta\gamma t_{1}}{2},
\nonumber\\
f_{6}&=&\sin \Delta m \Delta t \cos \Delta m t_{1} \cosh \frac{\Delta\gamma t_{1}}{2},
\nonumber\\
f_{7}&=&\cosh \frac{\Delta\gamma\Delta t}{2} \sin \Delta m t_{1} \cosh \frac{\Delta\gamma t_{1}}{2},
\nonumber\\
f_{8}&=&\cos \Delta m \Delta t \sin \Delta m t_{1} \cosh \frac{\Delta\gamma t_{1}}{2}.
\label{fi}
\eea

Defining also some combinations of the $f_{i}$, belonging to the different semileptonic decay outcomes, and separated for the imaginary and real parts of the CPT violating parameter the probability expression can be simplified even further with
\bea
f^{R}_{f \ol f} \equiv -f_{1} + f_{2} - f_{3} - f_{4}, &
\quad 
f^{I}_{f \ol f}\equiv - f_{5} - f_{6} - f_{7} - f_{8},
\nonumber \\
f^{R}_{f f}\equiv f_{1} + f_{2} - f_{3} + f_{4}, &
\quad 
f^{I}_{f f}\equiv f_{5} - f_{6} - f_{7} + f_{8}.
\nonumber\\
\label{fIR}
\eea

In this formalism of the decay rates, all information is accessible for the direction-dependent phenomenology. Some general tendencies can be noted. 

Decay rates into mode $F$ at time $t_1$, and its conjugate $\ol F$ at time $t_2$, and its reverse, have some terms appearing with the same sign while others with the opposite. The same sign terms are functions only of $\Delta t$ and are even functions of it. They indicate oscillations without spacetime-symmetry violations. The factors $k_{f \ol f}$ and $k_{\ol f f}$ carry information about direct symmetry violation in the amplitudes, but in the investigations here are assumed to be equal $k_{f \ol f}=k_{\ol f f}$.

The probabilities now change to the form to
\bea
P_{f \ol f} &=&
k_{f \ol f}
\bigl[
\cosh \frac{\Delta\gamma\Delta t}{2} + \cos \Delta m \Delta t
\nonumber\\
&&
\hbox{\hskip-20pt}
+ \Im \xi \sin \Delta m \Delta t + \Re \xi \sinh\frac{\Delta\gamma \Delta t}{2}  
\nonumber\\
&&
\hbox{\hskip-20pt}
+ \Re \Delta \xi \, f^{R}_{f \ol f} + \Im \Delta \xi \, f^{I}_{f \ol f}
\bigr],
\nonumber\\
&&
\hbox{\hskip-20pt}
\hspace{-0.5 cm}
P_{\ol f f} = P_{f \ol f} (\Delta \xi \rightarrow - \Delta \xi, \xi \rightarrow - \xi),
\label{probfIR1}
\eea
and
\bea
P_{f f} &=&
k_{f f}
\bigl[
\cosh \frac{\Delta\gamma\Delta t}{2} - \cos \Delta m \Delta t
\nonumber\\
&&
\hbox{\hskip-20pt}
+ \Im \Delta \xi \sin \Delta m \Delta t + \Re \Delta \xi \sinh\frac{\Delta\gamma \Delta t}{2}
\nonumber\\
&&
\hbox{\hskip-20pt} 
+ \Re \Delta \xi \, f^{R}_{f f} + \Im \Delta \xi \, f^{I}_{f f} 
\bigr],
\nonumber\\
&&
\hbox{\hskip-20pt}
P_{\ol f \ol f} = P_{f f} (\Delta \xi \rightarrow - \Delta \xi, k_{f f} \rightarrow k_{\ol f \ol f}).
\label{probfIR2}
\eea
To isolate the parameter $\xi$, decay rate asymmetries need to be set up. To do that one can use the parity properties of the trigonometric functions describing time developments or the asymmetries due to the time development differences between particle and antiparticle in the CPT violating scenario. The former was explored in Ref.\ \cite{NMPhKy01}. 

Some properties of the $f_{i}$ which facilitate the analysis of possible asymmetries of this nature are given in Table \ref{table2}.  
\begin{table}
\begin{center}
\tabcolsep8pt
\begin{tabular}{|c|c|c|c|c|c|c|}
\hline
$f_{i}$ & $P_{f \ol f}$ & $P_{\ol f f}$ & $P_{f f}$ & $P_{\ol f \ol f}$ & $\Delta t$ & $\Delta \gamma=0$ \\
\hline
\hline
$f_{1}$ & $-$ & $+$ & $+$ & $-$ & odd & 0 \\
\hline
$f_{2}$ & $+$ & $-$ & $+$ & $-$ & odd & 0 \\
\hline
$f_{3}$ & $-$ & $+$ & $-$ & $+$ & even & 0 \\
\hline
$f_{4}$ & $-$ & $+$ & $+$ & $-$ & even & 0 \\
\hline
$f_{5}$ & $-$ & $+$ & $-$ & $+$ & odd & - \\
\hline
$f_{6}$ & $-$ & $+$ & $+$ & $-$ & odd & 0 \\
\hline
$f_{7}$ & $-$ & $+$ & $-$ & $+$ & even & - \\
\hline
$f_{8}$ & $-$ & $+$ & $+$ & $-$ & even & - \\
\hline
\end{tabular}
\vskip -5pt
\caption{Properties of the time development functions $f_{i}$. }
\label{table2}
\end{center}
\vskip -10pt
\end{table}
Here no attention is given to mixing due to CP violation and so $W$ is considered equal to 1. Direct CPT or CP violation is also disregarded. This can, however, be probed with differences in integrated rates of opposite-sign and same-sign events. 

The focus here is to isolate information about $\xi$ and limit the components of $\Delta a_{\mu}$. That has to be done in two steps. First, the needed decay rate asymmetry has to be defined, containing information about $\xi$ and/or $\Delta \xi$. Second,  specifics of decay kinematics and orientation dependence of $\xi$ have to be analyzed to find methods of component-by-component analysis of the SME coefficient. This will be done in Sec.\ \ref{dirdep}.
 
Here two asymmetries are considered. The first one below is the difference of  $P_{f \ol f}$ and $P_{\ol f f}$ 
\bea
\fr{P_{f \ol f}\, -\, P_{\ol f f}}{P_{f \ol f}\, + \, P_{\ol f f}} &=& (\cosh \frac{\Delta\gamma\Delta t}{2} + \cos \Delta m \Delta t)^{-1}
\nonumber\\
&&
\hbox{\hskip-20pt}
\hspace{-1.8 cm}
\times
\bigl[
\Im \xi \sin \Delta m \Delta t \, + \, \Re \xi \sinh\frac{\Delta\gamma \Delta t}{2} 
\nonumber\\
&&
\hbox{\hskip-20pt}
\hspace{-1.8 cm} 
- \Re \Delta \xi \, f^{R}_{f \ol f} - \Im \Delta \xi \, f^{I}_{f \ol f}\bigr] . 
\label{asym1}
\eea
This contains information about $\Im \xi$, $\Re \xi$ as well as $\Re \Delta \xi$ and $\Im \Delta \xi$. 

The second asymmetry is the difference between rates of $P_{\ol f \ol f}$ and $P_{f f}$. This is only dependent on the differences of $\xi$, which in turn comes from differences in meson momenta 
\bea
\fr{P_{\ol f \ol f}\, -\, P_{f f}}{P_{\ol f \ol f}\, + \, P_{f f}} &=&(\cosh \frac{\Delta\gamma\Delta t}{2} - \cos \Delta m \Delta t)^{-1}
\nonumber\\
&&
\hbox{\hskip-20pt}
\hspace{-1.5 cm}
\times
\bigl[
\Im \Delta \xi  \left(\sin \Delta m \Delta t - f^{I}_{f f}\right) 
\nonumber\\
&&
\hbox{\hskip-20pt}
\hspace{-1.5 cm}
- \Re \Delta \xi \left(\sinh\frac{\Delta\gamma \Delta t}{2} - f^{R}_{f f} \right)
\bigr]. 
\label{asym2}
\eea
The smallness of the quantity $\Delta \xi$ limits the ability to make measurements in the neutral $B$ system. However, since oscillation time scales are comparable to the time for which the different momenta persist, it is worthwhile to investigate such effects. It is theoretically significant because it is the quantity that directly relates to a CPT violating mechanism rooted in a direction-dependent background. It also is the quantity that relates to the second-order decoherence effect discussed in Sec.\ \ref{decoh}.

To see more explicitly the oscillation time development, the decay rate asymmetries are also presented here with $\Delta \gamma$ very small, characteristic of the $B$ system. Measuring $\Delta \gamma$ is one way Belle II is hoped to contribute to $B$ system studies. However, these simplified forms facilitate function fitting and comparison with no violations.  
\bea
\fr{P_{f \ol f}\, -\, P_{\ol f f}}{P_{f \ol f}\, + \, P_{\ol f f}} &=& (1 + \cos \Delta m \Delta t)^{-1} 
\nonumber\\
&&
\hbox{\hskip-20pt}
\times
\bigl[
\Im \xi \sin \Delta m \Delta t
\nonumber\\
&&
\hbox{\hskip-20pt}
+ \Im \Delta \xi (\sin \Delta m t_{1} + \sin \Delta m (\Delta t + t_{1}))
\nonumber\\
&&
\hbox{\hskip-20pt}
+ \fr{\Delta \ga \Delta t}{2} (\Re \xi + \Re \Delta \xi \cos \Delta m t_{1})
\nonumber\\
&&
\hbox{\hskip-20pt}
+ \fr{\Delta \ga}{2}t_{1} (\cos \Delta m t_{1} + \cos \Delta m (\Delta t + t_{1}))
\bigr],
\nonumber\\
\fr{P_{\ol f \ol f}\, -\, P_{f f}}{P_{\ol f \ol f}\, + \, P_{f f}} &=& (1 - \cos \Delta m \Delta t)^{-1}
\nonumber\\
&&
\hbox{\hskip-20pt}
\times
\bigl[
\Im \Delta \xi \sin \Delta m \Delta t 
\nonumber\\
&&
\hbox{\hskip-20pt}
+ \Im \Delta \xi (\sin \Delta m \Delta t_{1} + \sin \Delta m (\Delta t-t_{1})) 
\nonumber\\
&&
\hbox{\hskip-20pt}
- \fr{\Delta \ga \Delta t}{2} \Re \Delta \xi \cos \Delta m t_{1} 
\nonumber\\
&&
\hbox{\hskip-20pt}
- \fr{\Delta \ga \Delta t_{1}}{2} (\cos \Delta m t_{1} - \cos \Delta m (\Delta t-t_{1}))
\bigr].
\nonumber\\
&&
\hbox{\hskip-20pt}
\label{asym2nogamma}
\eea
The above asymmetries can be expressed using only $\Im \xi$ and $\Im \Delta \xi$ with relation \rf{reim}.
In summary: the second asymmetry is only dependent on $ \Delta \xi$. Its nonzero value is a feature of momentum dependence. It would vanish to first order in the traditional approach, even if mass or decay rate difference exists between the particles, due to some explicit CPT breaking. 

Such phenomenon is included here in the terms containing the sum of the $\xi$'s, defined originally as the constant $\xi$ parameter. It appears in the first asymmetry, which can be nonzero to first order in both approaches. Traditionally, however, it would only vary with $\Delta t$. 

Both asymmetries show that the time development before the first decay influences the asymmetries. This influence always enters with $\Delta \xi$, and is a manifestation of different interaction of the two $B$ particles propagating in different directions relative to the background. 
\subsection{Direction dependence}
\label{dirdep}
\begin{figure}
\begin{center}
\includegraphics[width=3 in]{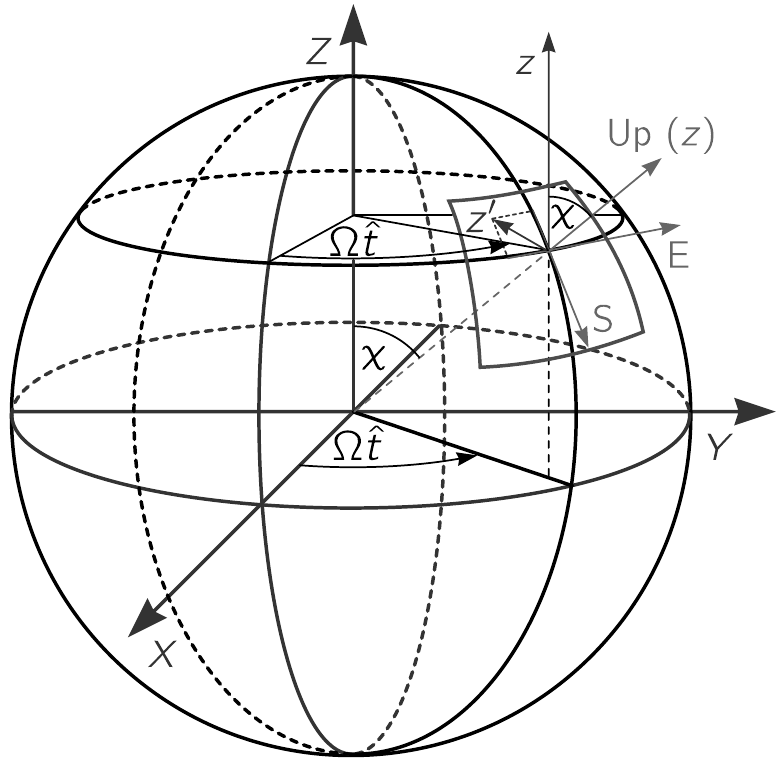}
\end{center}
\caption{Standard coordinate systems.}
\label{frames}
\end{figure} 
To take into account the direction dependence in the SME framework, two coordinate systems are defined. The standard frame used to describe the orientation of the $B$ momenta relative to the background as the Earth rotates is a Sun-centered frame. Its $\Z$ axis points to the celestial north pole at equinox 2000.0 at declination $90^{\circ}$. The standard $\X, \, \Y$ axes are at declination $0^{\circ}$, with right ascension $0^{\circ}$ for $\X$ and $90^{\circ}$ for $\Y$. In this frame $\De a_X,\De a_Y,\De a_Z$ are the three spatial components of $\De a_\mu$.

There is also a local laboratory coordinate system defined. In this paper the laboratory frame is named with vertical up ($\U$), East ($\E$) and South ($\S$) axes, corresponding to the natural geographic directions to form the $(\E,\S,\U)$ coordinate system. Both frames are shown in Fig.\ \ref{frames}. In the figure $\Om \hat t$ indicates the sidereal rotation of the lab frame. The colatitude $\ch$ of the detector is marked both in the $\X\Y\Z$ frame and the laboratory frame. $\U$ is shown as pointing along Earth's radius in the upward direction at the location of the detector. It precesses about $\Z$ with the Earth's sidereal frequency $\Om$. 

In the marked square section the intersection of a latitude and longitude circle defines the detector location. Axes $\E$ and $\S$ are defined tangential to these latitude and longitude lines, respectively. The figure also shows axis $z'$ indicating a beam direction in the $\E$-$\S$ plane. For detailed original description of the two coordinates systems see, for example, Ref.\ \cite{Mattcoords02}. 

In the directional analysis, what is sought is the sum and difference in $\xi$ of $B^0$ and $\overline{B^0}$, which is a velocity-dependent measure. However, experimentally, conservation of momentum is used in evaluating the detector kinematics, which detects momenta, rather than velocity. As was discussed in Sec.\ \ref{Mesfac}, in the SME the velocity vector is not necessarily parallel to the momentum, which was neglected in the indicated momentum dependencies due to it being second order in the SME coefficients. Here the differences in $\xi$ depend on small but macroscopically significant differences in particle-antiparticle momenta of the detector kinematics, which dominate over any difference in direction between velocity and momentum vectors in a Lorentz-violating background.
 
In the laboratory frame the general form of the $B$ meson velocity vector, $\vec\be$, is
\bea
\vec\be= \be (\sin\th\cos\ph \hat S + \sin\th\sin\ph \hat E + \cos\th \hat U),
\label{genbeta}
\eea 
where the angles $\th$ and $\ph$ are the conventional polar coordinates within the laboratory frame, with $\th$ being the angle to the $\U$ direction, and $\ph$ giving the direction East of South of the beam.
%The momentum magnitude is $p \equiv |\vec p| =\be m_B \ga(p)$. $\ga(p) = \sqrt{1 + p^2/m_B^2}$ as usual. 
The geographic directions connect to the Sun-centered frame as
\bea
\hat S &=& \cos\chi\cos\Omega \hat t\, \hat X + \cos\chi\sin\Omega \hat t\, \hat Y - \sin\chi\, \hat Z,
\nonumber\\
&&
\hbox{\hskip-20pt}
\hspace{-0.3 cm}
\hat E=\cos\Omega \hat t\, \hat Y - \sin\Omega \hat t\, \hat X,
\nonumber\\
&&
\hbox{\hskip-20pt}
\hspace{-0.3 cm}
\hat U= \sin\chi\cos\Omega\hat t\, \hat X + \sin\chi\sin\Omega\hat t\, \hat Y + \cos\chi\, \hat Z.
\label{SEU}
\eea
The most general expression for $\xi$ is given in Eq.\,(14) of Ref.\ \cite{NMPhKy01}.  
%\bea
%\xi &\equiv &
%\xi(\hat t, \vec p) \equiv \xi(\hat t, p, \th, \ph) 
%\nonumber\\
%&=& 
%\fr 
%{\ga}
%{\De \la} 
%\bigl\{
 %+ \be \De a_Z 
%(\cos\th\cos\ch - \sin\th \cos\ph\sin\ch)
%\nonumber\\
%&&
%\qquad
%+\be \bigl[
%\De a_Y (\cos\th\sin\ch 
%+ \sin\th\cos\ph\cos\ch )
%\nonumber\\
%&&
%\qquad\qquad
%-\De a_X \sin\th\sin\ph 
%\bigr] \sin\Om \hat t
%\nonumber\\
%&&
%\qquad
%+\be \bigl[
%\De a_X (\cos\th\sin\ch 
%+ \sin\th\cos\ph\cos\ch )
%\nonumber\\
%&&
%\qquad\qquad
%+\De a_Y \sin\th\sin\ph 
%\bigr] \cos\Om \hat t
%\bigr\}.
%\label{geom}
%\eea
%where $\hat t$ denotes the sidereal time.

Since the CPT violation depends on relative orientation to the background, detector location influences the measurement of the SME coefficient. Note, however, that components of $\Delta a_{\mu}$ lying in the equatorial plane are isolated by analyzing the sidereal rotation of Earth. This rotation and sidereal time dependence involved is the same for all detectors. It can be constrained for instance by sidereal binning. Here those components are the $\Delta a_{X}$ and $\Delta a_{Y}$ components. The time and $\Z$ components are often determined together, since neither has sidereal dependence. In what follows, the specific kinematics of the Belle II experiments is discussed with recommendations for appropriate analysis in the above context.

To understand better the geometry of the decay direction with respect to the constant background specific to Belle II, it is necessary to look at the form of $\xi$ and $\Delta \xi$. The sum of $\xi_{1}$ and $\xi_{2}$ in an asymmetric collider allows the constraining of all four components  of the SME coefficient while the difference of them can constrain the spatial ones. The equations below show that $\xi$ depends on the sum of the velocities $\beta^{\mu} \equiv \beta^{\mu}_{1} + \beta^{\mu}_{2}$ while $\Delta \xi$ on the difference $\Delta \beta^{\mu} \equiv \beta^{\mu}_{1} - \beta^{\mu}_{2}$. 

Observation of effects related to $\Delta \xi$ lead to information not only about CPT violation in general, but to a specific presentation and type of underlying mechanism of such violation. If the consistency of CPT violation with quantum field theory mandates momentum dependence, then experiments investigating $\Delta \xi$ carry fundamental importance. In observing it in $B$ decays, it amounts to looking at $\Delta \beta^{\mu}$. 

The $\Upsilon(4S)$ particle at Belle is boosted such that the $B^0$ and $\overline{B^0}$ are projected into a narrow cone. At the boost of $\beta\gamma = 0.28$, the maximum opening angle is only about $25^{\circ}$. 

\begin{figure}
\begin{center}
\includegraphics[width=3.2 in]{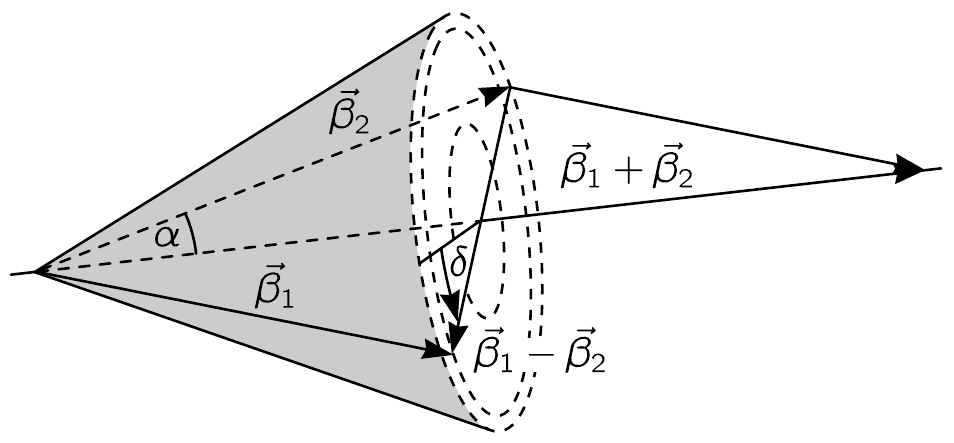}
\end{center}
\caption{Directional characteristics of $\vec {\beta}$ and $\De \vec {\beta}$ at Belle II.}
\label{Bdeccone}
\end{figure}

Figure\ \ref{Bdeccone} illustrates the situation when the angle between the two velocities $\vec{\beta_{1}}$ and $\vec{\beta_{2}}$ is maximumal and nearly equal. It shows the sum and the difference of the velocities, given explicitly as $\vec {\beta_{1}} + \vec {\beta_{2}}$, and $\vec {\beta_{1}} - \vec {\beta_{2}}$, respectively. 

Angle $\delta$ in the figure describes the rotation of $\Delta \vec {\beta} = \vec {\beta_{1}} - \vec {\beta_{2}}$ perpendicular to the horizontal plane containing the axis marked out by $\vec {\beta} = \vec {\beta_{1}} + \vec {\beta_{2}}$. Since the assumption here is that this plane coincides with the $\E$-$\S$ plane, compared to the definitions of Fig.\ \ref{frames}, $\delta$ is taken counterclockwise from the direction of West. 

In reality this distribution is valid only to a good approximation. Analyzing the kinematics, one also sees that one velocity vector subtends at most $3^{\circ}$ more than the other relative to the beam axis, so Fig.\ \ref{Bdeccone} represents closely the situation for any case. The only difference lies in the magnitude of $\Delta \vec {\beta}$. The sum of the velocities, $\vec {\beta}$, is nearly parallel to the beam axis while $\Delta \vec {\beta}$ is distributed in an approximately perpendicular circle around the beam. 

This simplifies the expressions for the sum and difference of the momentum. Expressed in the laboratory frame, $\vec {\beta}$ is fixed by the orientation of the beam at Belle II with known colatitude $\ch_{B}$. Assuming the beam is in the $E$-$S$ plane, there is no up component, and the East of South direction is at a given angle $\ph_{B}$. 

The only variation is in the sidereal rotation. In the expression below, fixed quantities carry index $B$. 
Based on the above description $\vec {\beta}$ and $\De \vec {\beta}$ are expressed in the laboratory frame by
\bea
\vec{\beta}&=&\beta(\cos\phi_{B}\hat S + \sin\phi_{B}\hat E),
\nonumber\\
&&
\hbox{\hskip-20pt}
\hspace{-0.3 cm}
\Delta \vec{\beta}= \Delta\beta(\sin\delta \hat U + \cos\delta\cos\phi_{B} \hat E - \cos\delta\sin\phi_{B} \hat S).
\nonumber\\
\label{bedebe}
\eea
According to expression \rf{xilambda} the dependence of $\xi$ and $\Delta \xi$ on $\beta$ and $\Delta \beta$ follows as
\bea
\xi &=&\xi_{1}+\xi_{2}=\fr{(\beta^{\mu}_{1} + \beta^{\mu}_{2})\Delta a_{\mu}}{\Delta \lambda},
\nonumber\\
&&
\hbox{\hskip-20pt}
\hspace{-0.3 cm}
\Delta \xi =\xi_{1}-\xi_{2}=\fr{(\beta^{\mu}_{1} - \beta^{\mu}_{2})\Delta a_{\mu}}{\Delta \lambda}.
\label{xideximom}
\eea
As discussed in Sec. \ref{SemlepDecrat}, $\Delta \xi$ plays an important role in uncovering the momentum dependent CPT violating phenomena. Expanding the second relation of Eq.\ \rf{xideximom} it takes the form 
\bea
\Delta \xi &=&
\nonumber\\
&&
\hbox{\hskip-20pt}
\frac{\gamma\Delta\beta}{\Delta\lambda}\bigl\{\bigl[(\sin\delta\sin\chi_{B}-\cos\delta\sin\phi_{B}\cos\chi_{B})\cos\Omega \hat t
\nonumber\\
&&
\hbox{\hskip-20pt}
-\cos\delta\cos\phi_{B}\sin\Omega\hat t \bigr]\Delta a_{X} +
\nonumber\\
&&
\hbox{\hskip-20pt}
\bigl[(\sin\delta\sin\chi_{B}-\cos\delta\sin\phi_{B}\cos\chi_{B})\sin\Omega \hat t
\nonumber\\
&&
\hbox{\hskip-20pt}
-\cos\delta\cos\phi_{B}\cos\Omega\hat t \bigr]\Delta a_{Y} +
\nonumber\\
&&
\hbox{\hskip-20pt}
\bigl[\sin\delta\cos\chi_{B}+\cos\delta\sin\phi_{B}\sin\chi_{B}\bigr]\Delta a_{Z}\bigr\}.
\label{dexigeom}
\eea

Inspecting expression \rf{dexigeom} suggests the detailed observation of both, sidereal time dependence as well angular distribution cylindrically around the detector. This can be done for instance by binning around the beam in $\delta$, as well as binning in sidereal time. Detector binning can lead to bounds on the $\Delta a_{Z}$ component, while sidereal binning can be used to give separate bounds on $\Delta a_{X}$ and $\Delta a_{Y}$. 

For $\xi$ in asymmetry \rf{asym1} the suitable form for Belle II using \rf{dexigeom} becomes 
\bea
\xi &=& \frac{\gamma}{\Delta \lambda}
\bigl\{
2 \Delta a_{0} +
\nonumber\\
&&
\hbox{\hskip-20pt}
\beta \bigl[(\cos\phi_{B}\cos\chi_{B}\cos\Omega \hat t - \sin\phi_{B}\sin\Omega \hat t)\Delta a_{X} +
\nonumber\\
&&
\hbox{\hskip-20pt}
(\cos\phi_{B}\cos\chi_{B}\sin\Omega \hat t - \sin\phi_{B}\cos\Omega \hat t)\Delta a_{Y} -
\nonumber\\
&&
\hbox{\hskip-20pt}
\cos\phi_{B}\sin\chi_{B}\Delta a_{Z}\bigr]
\bigr\}.
\label{xigeom}
\eea
Since this expression is not a function of $\delta$, here only the sidereal variation plays a role and binning data with Earth's rotation can give information about $\Delta a_{X}$ and $\Delta a_{Y}$. Information gained in Eq.\ \rf{dexigeom} can then help isolate $\Delta a_{0}$ from $\Delta a_{Z}$.
\subsection{Decoherence}
\label{decoh}
Any time a boosted particle interacts with a constant Lorentz-violating background, which remains unchanged under particle transformations, the phenomenological parameter of CPT violation becomes dependent on boost and orientation. 
Inspecting the progression of how such phenomenology effects the time evolution of the neutral mesons, one finds that the time evolution before the decay of the first particle has significance and a small probability for decoherence of the entangled states exists. 

It was already discussed that such effect occurs only due to the difference between the CPT violation parameter for different momenta and only to second order in $\Delta\xi$, a feature of the SME framework \cite{NMPhKy01,NMPhKy99,NMPhKy98}. In this framework it is also assumed that quark fields of different flavor couple to the background differently. Such effect, however, does not appear in the usual formalism that  assumes CPT violation only as a mass and/or decay rate difference of particle and antiparticle.  

Consider what would happen if the two particles would decay at the same time into an inclusive mode $F=l^{-}X$, producing same sign leptons. We expect this decay to occur with a vanishing amplitude, since by Bose statistics both $B$'s cannot oscillate into the same state at any given time. The last two equations of Eq.\ \rf{amplambda} give the amplitudes that describe decays into the same-flavor states.

Some properties were previously discussed and it was shown that to first order these give zero probability. Note that to go to second order, the original amplitude equation has to be evaluated. However, the extra term that appears in the full calculation does not produce any nonzero terms compared to using the expression below. The reason is that the relevant extra term has either a multiplying function that is zero for $\Delta t=0$ or produces higher than second order $\xi$ terms. There are two relevant amplitudes for coherence studies,
\bea
A_{f f} &=&
\frac{F^{2} W^{-1}}{2\sqrt{2}}
{k(t)}\left[\Delta\xi h_{2}(t)-\Delta\xi h_{2}(\Delta t) - 2 i h_{1}(\Delta t)\right],
\nonumber\\
\hbox{\hskip-20pt}
A_{\ol f \ol f} &=&
\frac{\ol F ^{2} W}{2\sqrt{2}}
{k(t)}\left[\Delta\xi h_{2}(t)-\Delta\xi h_{2}(\Delta t) + 2 i h_{1}(\Delta t)\right].
\nonumber\\
&&
\hbox{\hskip-20pt}
\label{amplambda2}
\eea
Take $W = 1$ and $F = \ol F$. For decays at the same time denote $t_{1} = t_{2} \equiv t_{D}$.
The amplitude squared to second order in $\Delta \xi$ for decays into same-flavor modes with $\Delta t=0$ is
\bea
P_{f f}&=& P_{\ol f \ol f} = 
\nonumber\\
&&
\hbox{\hskip-20pt}
\hspace{-1 cm}
\frac{1}{4}|F^{2}|^{2}|\Delta\xi |^{2} e^{-\gamma t_{D}}(\cosh\frac{\Delta\gamma t_{D}}{2} -\cos \Delta m t_{D})^{2}.
\label{decohprob}
\eea 

This gives specific means to constrain decoherence by placing a bound on $\De\xi$. The reverse is also true and limits on decoherence constrain the CPT-violating phenomenological parameter and through that $\Delta a_{\mu}$. Technically, tagging and reconstructing cannot be used if decoherence is a possibility, hence for specific studies of quantum correlations other methods must be applied.  

CPT violation in relation to entangled neutral mesons has also been studied in various quantum gravity scenarios and early universe models the possible loss of quantum coherence due to a topologically nontrivial spacetime. has been investigated and applied to studies of neutrinos and neutral mesons \cite{Mavro}. There are also more general theory-based searches proposed such as in \cite{Alok,Viennabeauty}. 

In the SME CPT violation comes from an effective hamiltonian that does not commute with the CPT operator due to the unequal diagonal elements resulting from spontaneous Lorentz violation. Disentanglement is a small effect that emerges naturally from the proper field theory that yields directional dependence. The searches described here take advantage of a thorough analysis of specific SME phenomenology to constrain that decoherence. There exists another approach leading to what is called the "$\omega$ effect", where the influence of a quantum gravity background leads to an ill-defined CPT operator and hence to an effective low-energy decoherence. In this case the entangled wave function is extended with a coherence weakening term \cite{novelcptv}. 

While the two frameworks differ, their experimental testing overlaps and can be used to gain information about SME based distanglement both in $K$ and $B$ factories. One of the best bounds come from KLOE where the $\omega$ parameter was constrained to $10^{-7}$ \cite{kaoncorrel,ambrosino}. It involves plotting the intensity for decays into the same final states as a function of the decay time difference scaled over lifetimes. This method is viable for SME-based tests as well.  While KLOE has good kinematic properties for studying this phenomenon due to its low boost, the assumed flavor and possible mass dependence of the SME formalism motivates searches in the other neutral meson systems as well. 

There are various proposals for investigations for the $B$ factory. Belle II can provide very high integrated luminosity. An observation of a significant number of decays into the same neutral $B$ state at the same time would be a clear signal of Planck-scale physics. A detailed discussion with focus on decay modes for a separate study of the T, CP and CPT symmetries, including CPT violation of the $\omega$ type and of the type involving unequal masses of particles and antiparticles, has been recently given for the $B$ system \cite{genuinecpt}. 

The approach presented in Ref. \cite{BelleEPR07} for searching for disentanglement in the $B$ system uses the raw asymmetry of opposite-flavor and same-flavor decay rates for a comparison study of the symmetry-breaking scenarios. $\Delta \gamma$ is assumed zero, so standard quantum mechanics would predict a $\cos \Delta m \Delta t$ function. Shifts from this function would be due to the terms containing $\Delta \xi$ in Eq.\ \rf{probfIR2}. Similar plots can be created using $\Delta t$ binning and detector binning with the decay rate asymmetry that specifically relates the decoherence to $\De\xi$. As is seen in Sec.  \ref{dirdep}, the $B$ decays have their particular distribution of the difference in momenta that $\De\xi$ depends on. These investigations are within the reach of the improved detector.
  
\section{Summary}
\label{sum}
This work has studied the phenomenology of correlated neutral meson decays within a general momentum-dependent formalism based on the SME framework of possible spontaneous CPT symmetry breaking. The phenomenology is suitable for any meson factory search, with a specific study focusing on possible experiments for the updated $B$ factory at Belle II. 

It is concluded that a component-by-component test of the relevant SME coefficient is feasible by studying appropriate decay rate asymmetries containing the sum and difference of the CPT-violating phenomenological parameter $\xi$. Appropriate bounds can be placed also using the kinematic properties of the decays. Earth geometry and kinematics are presented in detail. An understanding of momentum dependence and suitable binning of data facilitates the isolation of separate bounds on the components.

In the phenomenology, decoherence effects could only be ruled out to first order in this approach. Because of its theoretical significance of connecting the momentum dependence to nonzero asymmetries between same-flavor rates and to disentanglement, an investigation of quantum correlations was proposed. The analysis involves more detailed observation of same sign decay rates and specific study of same-flavor decays occurring at the same decay times. Based on detailed expansion of the time development functions to allow better fitting, keeping track of the influence the background has on the oscillation amplitudes, even before the decay of the first meson. 

The increase in integrated luminosity of Belle II is expected to give an order of magnitude raw increase in limits placed on the SME coefficient. However, improvements in data gathering and processing based on the more detailed phenomenology must be explored to give both qualitative and further quantitative improvement. Momentum and direction-dependent studies have a much better outlook due to better full event identification and vertex detection. While KLOE has excellent results constraining all aspects of this formalism, dependence on quark content of the SME coefficients encourages searches at the $B$ factory as well.  

\end{document}